# STATISTICAL MECHANICS OF NEOCORTICAL INTERACTIONS: EEG EIGENFUNCTIONS OF SHORT-TERM MEMORY

Lester Ingber

Lester Ingber Research
PO Box 06440
Wacker Dr PO Sears Tower
Chicago, IL 60606

and

DRW Investments LLC
311 S Wacker Dr Ste 900
Chicago, IL 60606

ingber@ingber.com, ingber@alumni.caltech.edu

ABSTRACT: This paper focuses on how bottom-up neocortical models can be developed into eigenfunction expansions of probability distributions appropriate to describe short-term memory in the context of scalp EEG. The mathematics of eigenfunctions are similar to the top-down eigenfunctions developed by Nunez, albeit they have different physical manifestations. The bottom-up eigenfunctions are at the local mesocolumnar scale, whereas the top-down eigenfunctions are at the global regional scale. However, as described in several joint papers, our approaches have regions of substantial overlap, and future studies may expand top-down eigenfunctions into the bottom-up eigenfunctions, yielding a model of scalp EEG that is ultimately expressed in terms of columnar states of neocortical processing of attention and short-term memory.

## 1. INTRODUCTION

### 1.1. Categorization of Experimental Spatial-Temporal EEG

Many reasonable theoretical studies of synaptic-like or neuron-like structures have attempted experimental confirmation. However, most often these investigations do not necessarily address the "spatial-temporal" scales they purport to describe. Many investigators would like to see more work on experimental design/tests of local-global interactions correlated to behavioral states at specific scales. For example, if an EEG could reasonably be correlated to a resolution of 3–5 cm within a time scale of 1–3 msec, then experiments should be attempted to test if specific states of attentional information-processing are highly correlated within this specific spatial-temporal range.

In this context, the work of Paul Nunez (PN) stresses resolution of EEG data within specific spatial-temporal scales, giving us candidate data for such correlations. It is most important for researchers to deal with the details of experimental evidence, not just pay homage to its existence.

### 1.2. Theoretical Descriptions of Spatial-Temporal EEG

The theoretical framework given by PN encompasses global and local neuronal columnar activity, giving a primary role to global activity. His work has generated interest in other investigators to take similar approaches to describing neocortical activity (Jirsa & Haken, 1996).

Other investigators, myself included, give a primary role to local activity, immersed in global circuitry. In this context, PN brings to BBS commentary a sound framework in which to further analyze the importance of considering multiple scales of neocortical activity.

### 1.3. Generality of Eigenfunction Expansion

Here, I add my own emphasis this subject, to address how eigenfunction expansions of models of brain function, similar to those performed by PN to describe how global models of wave phenomenon can be used effectively to describe EEG, can be applied to probability distributions of short-term memory fits to EEG data.

In the following description, emphasis is placed on overlap and collaboration with the work of PN, especially in areas wherein local and global interactions are required to detail models of neocortical interactions giving rise to EEG phenomena.

## 2. SMNI Description of Short-Term Memory (STM)

Since the early 1980's, a series of papers on the statistical mechanics of neocortical interactions (SMNI) has been developed to model columns and regions of neocortex, spanning mm to cm of tissue. Most of these papers have dealt explicitly with calculating properties of short-term memory (STM) and scalp EEG in order to test the basic formulation of this approach (Ingber, 1981; Ingber, 1982; Ingber, 1983; Ingber, 1984; Ingber, 1985a; Ingber, 1985b; Ingber, 1986; Ingber & Nunez, 1990; Ingber, 1991; Ingber, 1992; Ingber, 1994; Ingber & Nunez, 1995; Ingber, 1995a; Ingber, 1995b; Ingber, 1996; Ingber, 1997; Ingber, 1998). This model was the first physical application of a nonlinear multivariate calculus developed by other mathematical physicists in the late 1970's (Graham, 1977; Langouche $et$ $al$, 1982).

### 2.1. Statistical Aggregation

SMNI studies have detailed a physics of short-term memory and of (short-fiber contribution to) EEG phenomena (Ingber, 1984), in terms of $M^G$ firings, where $G$ represents $E$ or $I$, $M^E$ represents contributions to columnar firing from excitatory neurons, and $M^I$ represents contributions to columnar firing from inhibitory neurons. About 100 neurons comprise a minicolumn (twice that number in visual cortex); about 1000 minicolumns comprise a macrocolumn. A mesocolumn is developed by SMNI to reflect the convergence of short-ranged (as well as long-ranged) interactions of macrocolumnar input on minicolumnar structures, in terms of synaptic interactions taking place among neurons (about 10,000 synapses per neuron). The SMNI papers give more details on this derivation.

In this SMNI development, a Lagrangian is explicitly defined from a derived probability distribution of mesocolumnar firings in terms of the $M^G$ and electric potential variables, $\Phi^G$. A mechanical string model, as first discussed by PN as a simple analog of neocortical dynamics (Nunez, 1989; Nunez & Srinivasan, 1993), is derived explicitly for neocortical interactions using SMNI (Ingber & Nunez, 1990). In addition to providing overlap with current EEG paradigms, this defines a probability distribution of firing activity, which can be used to further investigate the existence of other nonlinear phenomena, e.g., bifurcations or chaotic behavior, in brain states.

### 2.2. STM

The SMNI calculations are of minicolumnar interactions among hundreds of neurons, within a macrocolumnar extent of hundreds of thousands of neurons. Such interactions take place on time scales of several $\tau$, where $\tau$ is on the order of 10 msec (of the order of time constants of cortical pyramidal cells). This also is the observed time scale of the dynamics of STM. SMNI hypothesizes that columnar interactions within and/or between regions containing many millions of neurons are responsible for these phenomena at time scales of several seconds. That is, the nonlinear evolution at finer temporal scales gives a base of support for the phenomena observed at the coarser temporal scales, e.g., by establishing mesoscopic attractors at many macrocolumnar spatial locations to process patterns in larger regions.

## 3. SMNI Description of EEG

### 3.1. EEG Regional Circuitry of STM Local Firings

Previous calculations of EEG phenomena (Ingber, 1985a), show that the short-fiber contribution to the $\alpha$ frequency and the movement of attention across the visual field are consistent with the assumption that the EEG physics is derived from an average over the fluctuations of the system. I.e., this is described by the Euler-Lagrange equations derived from the variational principle possessed by $L_\Phi$, which yields the string model described above (Ingber, 1982; Ingber, 1983; Ingber, 1988).

### 3.2. Individual EEG Data

The 1996 SMNI project used evoked potential (EP) EEG data from a multi-electrode array under a variety of conditions, collected at several centers in the United States, sponsored by the National Institute on Alcohol Abuse and Alcoholism (NIAAA) project (Zhang *et al*, 1995). The earlier SMNI 1991 study used only averaged EP data.

After fits were performed on a set of training data (Ingber, 1997), the parameters for each subject were used to generate CMI for out-of-sample testing data for each subject (Ingber, 1998). The results illustrate that the CMI give enhanced patterns to exhibit differences between the alcoholic and control groups of subjects.

## 4. EEG Eigenfunctions of STM

### 4.1. STM Eigenfunctions

The study fitting individual EEG data to SMNI parameters within STM-specific tasks can now be recast into eigenfunction expansions yielding orthogonal memory traces fit to individual EEG patterns. This development was described in the first SMNI papers.

The clearest picture that illustrates how this eigenfunction expansion is achieved is in (Ingber & Nunez, 1995) wherein, within a tenth of a second, there are stable multiple non-overlapping Gaussian-type peaks of an evolving probability distribution with the same STM constraints used in the NIH EEG study above. These peaks can be simply modeled as a set of orthogonal Hermite polynomials.

### 4.2. Expansion of Global EEG Eigenfunctions into Local STM Eigenfunctions

In the "classical" limit defined by the variational Euler-Lagrange equations described above, we have demonstrated how the local SMNI theory reduces to a string model similar to the global model of PN (Ingber & Nunez, 1990). One reasonable approach to developing the EEG eigenfunctions in the approach given by PN is to apply the variational derivatives directly to the SMNI STM Hermite polynomials, thereby yielding a model of scalp EEG that is ultimately expressed in terms of states of neocortical processing of attention and short-term memory.

## 5. CONCLUSION

PN has been a key exponent of realistic modeling of realistic neocortex for many years, even when it was not as popular as neural network modeling of "toy brains." In the process of insisting on dealing with aspects of models of neocortical systems that could be experimentally verified/negated, he has contributed to a rich approach to better understanding the nature of experimental EEG.

His approach has led him to stress appreciation of neocortex as functioning on multiple spatial-temporal scales, and he has collaborated with other investigators with different approaches to these multiple scales of interactions. His approach has been to formulate a top-down global model appropriate to the scale of scalp EEG phenomena, which includes some important local mesocolumnar features. This approach has been sufficiently robust to overlap with and enhance the understanding of other bottom-up approaches such as I have described, wherein mesocolumnar models appropriate to the scale of STM are scaled up to global regions appropriate to EEG.

In this paper, I have focussed on how bottom-up SMNI models can be developed into eigenfunction expansions of probability distributions appropriate to describe STM in the context of EEG. The mathematics of eigenfunctions are similar to the top-down eigenfunctions developed by PN, albeit they have different physical manifestations. The bottom-up eigenfunctions are at the local mesocolumnar scale, whereas the top-down eigenfunctions are at the global regional scale. Future studies may expand top-down eigenfunctions into the bottom-up eigenfunctions, yielding a model of scalp EEG that is ultimately expressed in terms of columnar states of neocortical processing of attention and short-term memory.

**REFERENCES**

Graham, R. (1977) Covariant formulation of non-equilibrium statistical thermodynamics. *Z. Physik B* 26:397-405.

Ingber, L. (1981) Towards a unified brain theory. *J. Social Biol. Struct.* 4:211-224.

Ingber, L. (1982) Statistical mechanics of neocortical interactions. I. Basic formulation. *Physica D* 5:83-107. [URL http://www.ingber.com/smni82_basic.ps.gz]

Ingber, L. (1983) Statistical mechanics of neocortical interactions. Dynamics of synaptic modification. *Phys. Rev. A* 28:395-416. [URL http://www.ingber.com/smni83_dynamics.ps.gz]

Ingber, L. (1984) Statistical mechanics of neocortical interactions. Derivation of short-term-memory capacity. *Phys. Rev. A* 29:3346-3358. [URL http://www.ingber.com/smni84_stm.ps.gz]

Ingber, L. (1985a) Statistical mechanics of neocortical interactions. EEG dispersion relations. *IEEE Trans. Biomed. Eng.* 32:91-94. [URL http://www.ingber.com/smni85_eeg.ps.gz]

Ingber, L. (1985b) Statistical mechanics of neocortical interactions: Stability and duration of the 7+−2 rule of short-term-memory capacity. *Phys. Rev. A* 31:1183-1186. [URL http://www.ingber.com/smni85_stm.ps.gz]

Ingber, L. (1986) Statistical mechanics of neocortical interactions. *Bull. Am. Phys. Soc.* 31:868.

Ingber, L. (1988) Mesoscales in neocortex and in command, control and communications ($C^3$) systems, In: *Systems with Learning and Memory Abilities: Proceedings, University of Paris 15-19 June 1987*, ed. J. Delacour & J.C.S. Levy. Elsevier, 387-409.

Ingber, L. (1991) Statistical mechanics of neocortical interactions: A scaling paradigm applied to electroencephalography. *Phys. Rev. A* 44:4017-4060. [URL http://www.ingber.com/smni91_eeg.ps.gz]

Ingber, L. (1992) Generic mesoscopic neural networks based on statistical mechanics of neocortical interactions. *Phys. Rev. A* 45:R2183-R2186. [URL http://www.ingber.com/smni92_mnn.ps.gz]

Ingber, L. (1994) Statistical mechanics of neocortical interactions: Path-integral evolution of short-term memory. *Phys. Rev. E* 49:4652-4664. [URL http://www.ingber.com/smni94_stm.ps.gz]

Ingber, L. (1995a) Statistical mechanics of multiple scales of neocortical interactions, In: *Neocortical Dynamics and Human EEG Rhythms*, ed. P.L. Nunez. Oxford University Press, 628-681. [ISBN 0-19-505728-7. URL http://www.ingber.com/smni95_scales.ps.gz]

Ingber, L. (1995b) Statistical mechanics of neocortical interactions: Constraints on 40 Hz models of short-term memory. *Phys. Rev. E* 52:4561-4563. [URL http://www.ingber.com/smni95_stm40hz.ps.gz]

Ingber, L. (1996) Statistical mechanics of neocortical interactions: Multiple scales of EEG, In: *Frontier Science in EEG: Continuous Waveform Analysis (Electroencephal. clin. Neurophysiol. Suppl. 45)*, ed. R.M. Dasheiff & D.J. Vincent. Elsevier, 79-112. [Invited talk to Frontier Science in EEG Symposium, New Orleans, 9 Oct 1993. ISBN 0-444-82429-4. URL http://www.ingber.com/smni96_eeg.ps.gz]

Ingber, L. (1997) Statistical mechanics of neocortical interactions: Applications of canonical momenta indicators to electroencephalography. *Phys. Rev. E* 55:4578-4593. [URL http://www.ingber.com/smni97_cmi.ps.gz]

Ingber, L. (1998) Statistical mechanics of neocortical interactions: Training and testing canonical momenta indicators of EEG. *Mathl. Computer Modelling* 27:33-64. [URL http://www.ingber.com/smni98_cmi_test.ps.gz]

Ingber, L. & Nunez, P.L. (1990) Multiple scales of statistical physics of neocortex: Application to electroencephalography. *Mathl. Comput. Modelling* 13:83-95.

Ingber, L. & Nunez, P.L. (1995) Statistical mechanics of neocortical interactions: High resolution path-integral calculation of short-term memory. *Phys. Rev. E* 51:5074-5083. [URL http://www.ingber.com/smni95_stm.ps.gz]

Jirsa, V.K. & Haken, H. (1996) Field theory of electromagnetic brain activity. *Phys. Rev. Lett.* 77:960-963.

Langouche, F., Roekaerts, D. & Tirapegui, E. (1982) *Functional Integration and Semiclassical Expansions*. Reidel, Dordrecht, The Netherlands.

Nunez, P.L. (1989) Towards a physics of neocortex, Vol. 2, In: *Advanced Methods of Physiological System Modeling*, ed. V.Z. Marmarelis. Plenum, 241-259.

Nunez, P.L. & Srinivasan, R. (1993) Implications of recording strategy for estimates of neocortical dynamics with electroencephalography. *Chaos* 3:257-266.

Zhang, X.L., Begleiter, H., Porjesz, B., Wang, W. & Litke, A. (1995) Event related potentials during object recognition tasks. *Brain Res. Bull.* 38:531-538.